\documentclass[aps,prb,twocolumn,showpacs,preprintnumbers,amsmath,amssymb,superscriptaddress]{revtex4}%

\usepackage{graphicx}
\usepackage{dcolumn}
\usepackage{bm}
\usepackage{color}

\begin{document}

\preprint{PREPRINT (\today)}

\newpage

\title{Tuning the superconducting and magnetic properties in Fe$_{y}$Se$_{0.25}$Te$_{0.75}$ by varying the Fe-content}

\author{M.~Bendele}
\email{markus.bendele@physik.uzh.ch}
\affiliation{Physik-Institut der Universit\"{a}t Z\"{u}rich, Winterthurerstrasse 190, CH-8057 Z\"{u}rich, Switzerland}
\affiliation{Laboratory for Muon Spin Spectroscopy, Paul Scherrer Institut, CH-5232 Villigen PSI, Switzerland}
\author{P.~Babkevich}
\affiliation{Department of Physics, Clarendon Laboratory, Oxford University, Oxford OX1 3PU, United Kingdom}
\affiliation{Laboratory for Neutron Scattering, Paul Scherrer Institut, CH-5232 Villigen PSI, Switzerland}
\author{S.~Katrych}
\affiliation{Laboratory for Solid State Physics, ETH Zurich, CH-8093 Z\"urich, Switzerland}

\author{S.~N.~Gvasaliya}
\affiliation{Laboratory for Neutron Scattering, Paul Scherrer Institut, CH-5232 Villigen PSI, Switzerland}

\author{E.~Pomjakushina}
\affiliation{Laboratory for Developments and Methods, Paul Scherrer Institut, CH-5232 Villigen PSI, Switzerland}

\author{K.~Conder}
\affiliation{Laboratory for Developments and Methods, Paul Scherrer Institut, CH-5232 Villigen PSI, Switzerland}

\author{B.~Roessli}
\affiliation{Laboratory for Neutron Scattering, Paul Scherrer Institut, CH-5232 Villigen PSI, Switzerland}

\author{A.~T.~Boothroyd}
\affiliation{Department of Physics, Clarendon Laboratory, Oxford University, Oxford OX1 3PU, United Kingdom}

\author{R.~Khasanov}
\affiliation{Laboratory for Muon Spin Spectroscopy, Paul Scherrer Institut, CH-5232 Villigen PSI, Switzerland}

\author{H.~Keller}
\affiliation{Physik-Institut der Universit\"{a}t Z\"{u}rich, Winterthurerstrasse 190, CH-8057 Z\"{u}rich, Switzerland}

\begin{abstract}
The superconducting and magnetic properties of Fe$_{y}$Se$_{0.25}$Te$_{0.75}$ single crystals ($0.9\leq y \leq1.1$) were studied by means of x-ray diffraction, SQUID magnetometry, muon spin rotation, and elastic neutron diffraction. The samples with $y<1$ exhibit coexistence of bulk superconductivity and incommensurate magnetism.  The magnetic order remains incommensurate for $y\geq 1$, but with increasing Fe content superconductivity is suppressed and the magnetic correlation length increases. The results show that the superconducting and the magnetic properties of the Fe$_{y}$Se$_{1-x}$Te$_{x}$ can be tuned not only by varying the Se/Te ratio but also by changing the Fe content.
\end{abstract}

\pacs{0 0 0}

\maketitle

The iron chalcogenide family of superconductors Fe$_{y}Ch$ ($Ch=$~Se/Te) was discovered in 2008\cite{Hsu08} shortly after the first report of high temperature superconductivity (HTS) in the iron pnictides.\cite{Kamihara08} This family stands out because of its simple crystal structure relative to other Fe-based superconductors.\cite{Rotter08,Wang08,Ogino09,Zhu09} In common with the other Fe-based HTS the parent phase, FeTe, exhibits antiferromagnetic order,\cite{Fruchart75,Bao09,Li09} and superconductivity appears only upon substitution of Te with Se or S.\cite{Yeh08,Mizugushi09} The superconducting transition temperature $T_{\rm c}$ is lower than in most of the other Fe-based superconductors, reaching a value of $\simeq 14\,$K in Fe$_{y}$Se$_{1-x}$Te$_{x}$ at optimal Se-Te ratio and $\simeq 36\,$K at high pressures.\cite{Hsu08,Khasanov09_FeSeTe,Margadonna09,Sales09} Despite its relatively low $T_{\rm c}$, the binary Fe$Ch$ system is attractive for fundamental investigations of the interplay between magnetism and superconductivity because of (i) its simple crystallographic structure, (ii) the relative ease with which single crystals can be grown, and (iii) the similarity of the Fermi-surface topology with that of other Fe-based superconductors.\cite{Subedi08}

Recently, it has been reported that the superconducting and magnetic properties of Fe$_{y}$Se$_{x}$Te$_{1-x}$ not only depend on the Se-Te ratio but also strongly on the Fe content.\cite{Viennois10,Zhang09,Liu09,Wen09} Here we report a systematic investigation of the magnetic and superconducting properties, and their interplay, of Fe$_{y}$Se$_{0.25}$Te$_{0.75}$ with different nominal Fe content in the range $0.9\leq y \leq1.1$. Samples with low Fe content ($y<1$) are found to be bulk superconductors with coexistent magnetic order that sets in at a temperature below $T_{\rm c}$. Stoichiometric samples ($y\sim1$) show filamentary superconductivity and magnetic order. Fe-rich samples ($y>1$) are almost purely magnetic with only traces of superconductivity. Interestingly, the magnetic order was observed to be incommensurate\cite{Khasanov09_FeSeTe,Babkevich10} throughout the entire range of nominal Fe content investigated, although the correlation length increases with increasing Fe content.

The Fe$_y$Se$_x$Te$_{1-x}$ samples were prepared within a wide range of nominal Fe content from $y=0.9$ to 1.1 ($y=0.90$, 0.95, 0.98, 1.00, 1.01, 1.02, 1.03, 1.07, 1.10), with a fixed Se to Te ratio of $x=0.25$. The single crystals were prepared in the form of rods with masses of $m\sim4-5\,$g by a modified Bridgman method. For a detailed description of the procedure see Ref. \onlinecite{Bendele10_FeSeTe}.

To establish the crystal structure and the stoichiometry, the samples were investigated by single-crystal X-ray diffraction (XRD) at room temperature. Data reduction and numerical absorption correction were performed using the Bruker AXS Inc software package.\cite{XRD1} The crystal structure was solved by direct method and refined on $F^2$, employing the programs SHELXS-97 and SHELXL-97.\cite{XRD2} All crystals reveal a tetragonal lattice (space group $P4/nmm$) with the lattice parameters $a$ and $c$ presented in Table \ref{table_xrd}. The refined Se/Te ratio Se$_{\rm occ}$ is within standard deviation ($\pm 3$\,\%) close to the nominal content. The three samples with the nominal Fe content of 0.95, 0.98, and 1.03 reveal 100\,\% ($\pm 2$\,\%) Fe occupation Fe$_{\rm occ}$ at the $2b$ site [($1/4,3/4,1/2$) for the space-group $P4/nmm$, origin choice 2]. The sample with $y=1.07$ shows an occupation of 100\,\% Fe at the $2b$ position and some sharp maximum (2.7\,\AA\ from the Se/Te atom) on a difference Fourier map $F_0 - F_{\rm c}$ indicating that the remaining Fe (8\,\%) partially occupies the interstitial $2c$ site ($3/4,3/4,0.2105(2)$).

\begin{figure}[t!]
\centering
\vspace{-0cm}
\includegraphics[width=1\linewidth]{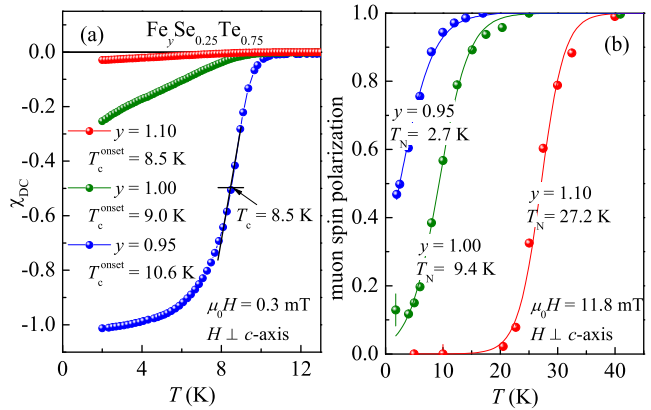}
\vspace{-1cm}
\caption{(color online) (a) Temperature dependence of the volume susceptibility $\chi_{\rm DC}$ of representative compositions ($y=0.95$, 1.00, and 1.10) of single-crystal Fe$_y$Se$_{0.25}$Te$_{0.75}$. The onset of the superconducting transition $T_{\rm c}^{\rm onset}$ and the midpoint corresponding to $\chi_{\rm DC}=-0.5$ are indicated. (b) Temperature dependence of the muon-spin polarization of the slow relaxing component ($P_{\rm slow}^{TF}$). The magnetic transition $T_N$ is determined from a fit to a Fermi-type function. }
\label{fig_magn_musr}
\end{figure}

\begin{table}[t!]\centering\caption{Summary of structural parameters obtained from single-crystal XRD of selected compositions of Fe$_y$Se$_{0.25}$Te$_{0.75}$. $h_{\rm Se/Te}$ denotes the height of $Ch$ above the Fe plane.}\hspace{20mm}
\begin{tabular}{l|c|c|c|c}\toprule
 &$y=0.95$ & $y=0.98$ & $y=1.03$ & $y=1.07$\\\hline
$a$-axis (\AA)	& 3.8125(5)	& 3.8096(3)	& 3.8090(4)	& 3.8104(3)	\\
$c$-axis (\AA)	& 6.1699(13)	& 6.1524(6)	& 6.1562(8)	& 6.1717(10)	\\
$h_{\rm Se/Te}$(\AA)& 1.717(2)	& 1.709(1)	& 1.715(1)	& 1.733(1)	\\
Fe$_{\rm occ}$	& 1.00(2)	& 1.00(2)	& 1.00(2)	& $1.00+0.08$\footnote{
1.00 at the $2b$ and 0.08 at the $2c$ site}	\\
Se$_{\rm occ}$	& 0.25(3)	& 0.26(3)	& 0.24(3)	& 0.23(3)	\\
\toprule
\end{tabular}
\label{table_xrd}
\end{table}
The superconducting properties of Fe$_{y}$Se$_{0.25}$Te$_{0.75}$ were studied on plate-like samples with a typical mass $\sim 50\,$mg that were always mounted with the flat surface ($ab$-plane) parallel to the magnetic field to minimize the demagnetization effect on the magnetic moment. The Zero-field cooled (ZFC) susceptibility measurements were performed with a Quantum Design 7\,Tesla Magnetic Property Measurement System (MPMS-XL7) SQUID Magnetometer in a magnetic field of $\mu_0 H=0.3\,$mT using the Reciprocating Sample Option (RSO). The data are shown in Fig.~\ref{fig_magn_musr}(a) for samples with representative doping. The measurements indicate that Fe$_{y}$Se$_{0.25}$Te$_{0.75}$ starting from a nominal Fe content of $y=0.90$ up to $y=0.98$ exhibits bulk superconductivity, since $\chi_{\rm DC}(2\, \rm{K})\simeq -1$ close to ideal diamagnetism. However, only the samples with the lowest Fe content ($y=0.90$ and $0.95$) show a saturation of the magnetic moment to $\chi\simeq-1$ expected for a Meissner state. Already for $y=0.98$ the transition is broad and tends to saturate only below $2\,$K. The onset of the superconducting transition decreases with increasing nominal Fe content: $T_{\rm c}^{\rm onset}\simeq11.5\,$K, $\simeq10.6\,$K and $\simeq8\,$K for $y=0.90$, 0.95 and 0.98. The samples with a nominal Fe content higher than $y\simeq1$ show only traces of superconductivity with a superconducting volume fraction of less than $\approx 30\,$\% at low temperatures for $y\sim 1.00$ and less than $\approx 5\,$\% for $y\geq 1.03$. The onset of the superconducting transition of the samples showing traces of supercondutivity is always at $T_{\rm c}^{\rm onset}\approx 9$~K. In order to distinguish samples with bulk superconductivity, $T_{\rm c}$ was defined as the midpoint of the superconducting transition (namely $\chi_{\rm DC}=-0.5$). Thus $T_{\rm c}\simeq9.7$, 8.5, and 6.5\,K for the compositions $y=0.90$, 0.95, and 0.98.

The magnetic response of the samples was investigated by Zero-field (ZF), transverse field (TF) and longitudinal field (LF) muon spin rotation ($\mu$SR) experiments carried out at the $\pi$M3 beam line at S$\mu$S at the Paul Scherrer Institute (PSI), Switzerland. In TF geometry the muons stopping in magnetic parts of the samples lose their polarization relatively fast, because the field at the muon stopping site is a superposition of the internal field and the applied external field of $11.8\,$mT. The internal field distribution is examined in ZF measurements, whereas LF experiments provide information whether the internal field is static or dynamic (fluctuating).

The ZF time spectra (not shown) of the bulk superconducting samples ($y=0.90$ and 0.95) show no difference between $T=20$\,K and $7\,$K. However, at lower temperatures an additional fast drop of the muon spin polarization $P(t)$ develops, and the $\mu$SR time spectra at $1.8\,$K and $20\,$K do not coincide any more. This suggests that magnetic ordering observed by $\mu$SR develops in the sample at temperatures below $T_c$. The data were described using the function:
\begin{equation}
 P^{\rm ZF}(t)=P^{\rm ZF}_{\rm fast}(0)e^{-\Lambda^{\rm ZF}_{\rm fast}t}+P^{\rm ZF}_{\rm slow}(0)e^{-\Lambda^{\rm ZF}_{\rm slow}t}
 \label{eq_ZF}
\end{equation}
Here, $P^{\rm ZF}_{\rm fast (slow)}$ and $\Lambda^{\rm ZF}_{\rm fast (slow)}$ are the initial ZF muon spin polarization and the exponential depolarization rate of the fast (slow) relaxing component, respectively. Upon increasing the nominal Fe content to $y\leq 0.98$ the change in the relaxation occurs at temperatures above $T_{\rm c}$.
At higher temperatures ($T\sim 130\,$K) an additional change of the ZF depolarization was observed, whose origin requires further investigation.

The magnetic ordering temperature was investigated by means of TF $\mu$SR [see Fig.~1(b)]. The TF time spectra (not shown) can be divided into a fast drop of the muon spin polarization within the first 100\,ns and a slow relaxing part below the temperature where magnetism starts to develop. Accordingly, the signal was divided into two parts:
\begin{eqnarray}
 \nonumber P^{\rm TF}(t)=P_{\rm fast}^{\rm TF}(0)\exp[-\Lambda^{\rm TF}_{\rm fast} t]\cos(\gamma_\mu B t+\phi)\\
         + P_{\rm slow}^{\rm TF}(0)\exp[-\Lambda^{\rm TF}_{\rm slow} t]\cos(\gamma_\mu B t+\phi)
 \label{eq_TF},
\end{eqnarray}
where $\gamma_\mu/2\pi=135.5\,$MHz/T is the muon gyromagnetic ratio, $\phi$ is the initial phase of the muon ensemble, and $\Lambda_{\rm fast (slow)}$ the exponential relaxation rate. The fast relaxing component $P_{\rm fast}$, attributed to the development of magnetism, represents the magnetic volume fraction of the sample and increases with decreasing temperature [Fig.~1(b)]. In the bulk superconducting samples ($y=0.90$ and 0.95) it occupies at the lowest investigated temperature ($T=1.6\,$K) more than $60\,$\% of the signal, indicating that $\geq 60\,$\% of the sample is magnetically ordered. Furthermore, the relaxation of the slow relaxing part increases just below $T_{\rm c}$ indicating the formation of a vortex lattice in the superconducting state. Upon increasing the nominal Fe content to $y\leq 0.98$ the samples were found to be $100\,$\% magnetic at $1.6\,$K as the muon spin polarization drops to zero within the first $100\,$ns. 

LF measurements reveal that the magnetic order is static in the bulk superconducting samples since the muon spin polarization recovers almost $100\,$\% at $B^{\rm LF}=0.64\,$T and the muon spins decouple from the static internal fields $B_{\rm int}$ at $B^{\rm LF}\simeq 10\cdot B_{\rm int}$.\cite{Schenk86} Thus the static internal field in the superconducting samples is $B_{\rm int}\gtrsim 0.1\,$T at the muon stopping site.

The ordering temperature $T_{\rm N}$ in Fig.~1(b) was determined by fitting a Fermi-type function $f(T)=[1+\exp( \beta(T_{\rm N}-T))]^{-1}$ ($\beta^{-1}$ is the with of transition) to the data [solid lines in Fig.~1(b)].\cite{Khasanov08_isotope} It develops with increasing nominal Fe content from $T_{\rm N}=1.7\,$K at the lowest Fe content ($y=0.90$) to 2.7\,K, 5\,K, 10\,K, 12.5\,K, and 17\,K for $y=0.95$, 0.98, 1.00, 1.01, and 1.02 respectively. Above $y=1.03$ it seems to saturate at $T_{\rm N}=27$\,K (see \onlinecite{Khasanov09_FeSeTe}), 28\,K and 29\,K for $y=1.03$, 1.07, and 1.10.  

\begin{figure}[t!]
\centering
\vspace{-0cm}
\includegraphics[width=1\linewidth]{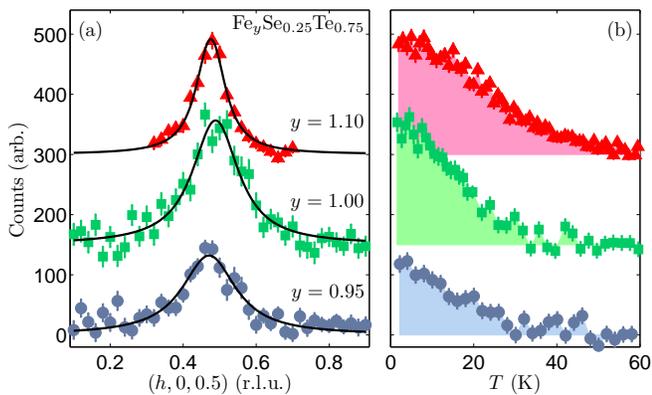}
\caption{(color online) Neutron diffraction from Fe$_{y}$Se$_{0.25}$Te$_{0.75}$ samples for $y =  0.95$, 1.00 and 1.10. (a) Intensity profiles along the $(h,0,0.5)$ direction at $T = 2$\,K  after subtraction of a background signal measured at $T\approx$50\,K. The solid line shows a fit to a Lorentzian function convoluted with the resolution function of the instrument. (b) Temperature dependence of the intensity of the magnetic reflection. In both panels the $y=1$ and $y=1.1$ data are displaced vertically for clarity.}
\label{fig_neutron}
\end{figure}

To investigate the change in the magnetic correlations with Fe content, we performed neutron diffraction measurements on single crystals of Fe$_{y}$Se$_{0.25}$Te$_{0.75}$ for $y = 0.95$, 1.00, and 1.10 at 1.5\,K. The experiments were performed on the triple-axis spectrometer TASP\cite{TASP} at the SINQ spallation source (PSI).\cite{SINQ} The instrument was configured for a neutron wave vector of 2.66\,\AA$^{-1}$ with no collimation. Each sample was aligned on nuclear Bragg reflections to an accuracy of better than 0.008\,r.l.u. at 1.5\,K.
Figure~\ref{fig_neutron}(a) shows elastic-scans in the $(h,0,l)$ scattering plane. A background recorded at around 50\,K (above the magnetic ordering temperature) was subtracted in order to isolate the magnetic contribution to the scattering at 1.5\,K. Diffuse incommensurate magnetic peaks centered at ${\bf q} = (1/2 - \delta,0,1/2)$, with $\delta \approx 0.03$, are observed in all three samples. These results together with our previous neutron diffraction studies show that samples in the entire range $0.95<y<1.10$ have incommensurate magnetic order.\cite{Khasanov09_FeSeTe,Babkevich10} We note that our results hint at a possible reduction in incommensurability $\delta$ for the $y=1.00$ sample. However, due to the broad nature of the peak this shift may not be related to the sample. As can be seen from Fig.~\ref{fig_neutron}(a), the magnetic peaks along $(h,0,0.5)$ appear to become broader with reduced Fe-content with correlation lengths along $a$ deduced to be 7.1(5)\,\AA, 8.4(6)\,\AA, and 13.8(8)\,\AA\ for $y = 0.95$, 1.00, and 1.10 respectively. This would suggest that the magnetic correlations become more short-ranged on lowering the Fe content. 

\begin{figure}[b!]
\centering
\vspace{-0cm}
\includegraphics[width=1\linewidth]{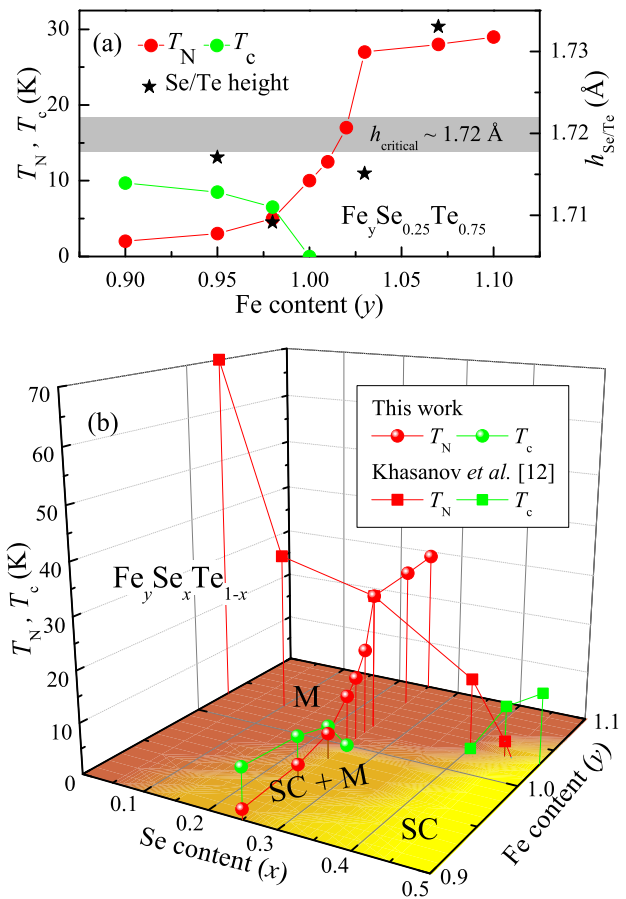}
\caption{(color online) (a) Variation of $T_{\rm c}$, $T_{\rm N}$ and the Se/Te height $h_{\rm Se/Te}$ in Fe$_y$Se$_{0.25}$Te$_{0.75}$ as a function of $y$. (b) Three dimensional phase diagram of $T_{\rm c}$ and $T_{\rm N}$ of Fe$_y$Se$_{x}$Te$_{1-x}$ as a function of $x$ and $y$. M and SC denote the magnetic and superconducting phases, respectively. The circles are from this work, and the squares are data taken from Khasanov {\it et al.}\cite{Khasanov09_FeSeTe}}
\label{fig_Phase_Diagram}
\end{figure}

The temperature at which magnetic order sets in appears also to be dependent on $y$ in Fe$_{y}$Se$_{0.25}$Te$_{0.75}$ as shown in Fig.~\ref{fig_neutron}(b). In the $y=1.10$ sample magnetic order is found to develop below $\approx 50$\,K. However, for $y=1.00$ and $y=0.95$ the transition temperature is reduced to around 30\,K. These results are in good agreement with the change in $T_{\rm N}$ with doping observed by $\mu$SR, but due to the difference in fluctuation rates sampled by neutrons and muons, the temperatures at which spin freezing occurs are not the same.

The results of the magnetization and $\mu$SR measurements of Fe$_{y}$Se$_{0.25}$Te$_{0.75}$ are summarized in a phase diagram in Fig.~\ref{fig_Phase_Diagram}(a). 
In the region of $y <1$ a coexistence of superconductivity and magnetism is observed. A recent nuclear magnetic resonance (NMR) study on BaFe$_{2-x}$Co$_x$As$_2$ showed the appearance of magnetic order on all Fe sites and ruled out nanoscale segregation in this material.\cite{Laplace09} More reasonable is a coexistence of the two order parameters on atomic scale, as was already suggested earlier \cite{Khasanov09_FeSeTe,Bendele10,Laplace09} and in agreement with models recently proposed.\cite{Vorontsov09,Cvetkovic09}

It has been predicted theoretically that the height of the chalcogenide above the Fe-plane $h_{\rm Se/Te}$ can influence the magnetic ordering in Fe$_y$Se$_{x}$Te$_{1-x}$.\cite{Moon10} For that reason $h_{\rm Se/Te}$ derived from XRD measurements is plotted in addition to $T_{\rm c}$ and $T_{\rm N}$ in Fig.~\ref{fig_Phase_Diagram}(a). The crossover from the purely magnetic to the coexistence region occurs at $h_{\rm Se/Te}\simeq1.72\,$\AA. This result is in good agreement with Density-Functional Theory (DFT) calculations, which predict a change from double- to single-stripe AFM ordering at $h_{\rm Se/Te}\sim 1.71-1.72\,$\AA.\cite{Moon10} However, the theoretical calculations are based on Se substitution only and do not take into account any excess Fe. Furthermore, $h_{\rm Se/Te}$ might not be the only factor since bulk superconductivity appears only for $y\leq0.98$ whereas the proposed critical height $h_{\rm Se/Te}=1.72$\,\AA\ is reached already at $y=1.03$.

The amount of excess Fe seems to play a major role in this system as only for $y\leq1$ does bulk superconductivity occur. Here the magnetic correlations in the system become more short-ranged and lead to less well correlated magnetic order as compared with $y>1$. Thus, only upon reducing the magnetic correlations by lowering the amount of Fe in the system does it become superconducting. Nevertheless, it seems unlikely that the excess Fe acts as isolated magnetic moments that destroy superconductivity. It might on the other hand act as a magnetic electron donor\cite{Zhang09} that suppresses superconductivity and induces weakly localized electronic states.\cite{Wen09,Liu09}

The tentative three-dimensional phase diagram of the transition temperatures $T_{\rm c}$ and $T_{\rm N}$ of Fe$_y$Se$_{x}$Te$_{1-x}$ for $0\leq x\leq 0.5$ and $0.9\leq y\leq 1.1$ is shown in Fig. \ref{fig_Phase_Diagram}(b). Fe$_y$Te is always antiferromagnetically ordered.\cite{Bao09,Li09} Upon substituting Te by Se the order becomes weaker while superconductivity is enhanced, and finally the system becomes bulk superconducting. This behavior can be tuned not only by the substitution of Se but also by adjusting the Fe content. The superconductivity is suggested to be of multi-band nature, where different doping channels might be involved.\cite{Bendele10_FeSeTe}

To conclude, we have found that the phase diagram of Fe$_y$Se$_{0.25}$Te$_{0.75}$ in the range $0.9\leq y \leq 1.1$ exhibits a strong dependence of its superconducting and magnetic phases on $y$. In the low Fe content region $y\leq 1$ bulk superconductivity and incommensurate magnetism coexist. With increasing $y$ the magnetic order becomes correlated over a longer range and superconductivity vanishes.  This work emphasizes that not only the Se/Te ratio, but also the Fe content is important in controlling the magnetic and superconducting properties of the iron chalcogenides. 

This work was performed at the S$\mu$S and SINQ, Paul Scherrer Institute (PSI, Switzerland). It was partially supported by the Swiss National Foundation and the NCCR program MaNEP.

\end{document}